\documentclass[11pt]{article}
\usepackage{color}
\usepackage{subfigure}
\usepackage{amsmath}
\usepackage{amssymb}
\usepackage{graphicx,color}
\usepackage{cite}
\usepackage{enumerate}
\usepackage{amsthm}
\usepackage{amsfonts,mathrsfs}
\usepackage{geometry}
\usepackage{ifthen}
\usepackage{graphicx}
\usepackage{epstopdf}
\usepackage{float}
\usepackage{bm}
\usepackage[caption=false]{subfig}
\usepackage[utf8]{inputenc}

\begin{document}

\title{\bf Coherence and entanglement dynamics in Shor's algorithm}

\vskip0.1in
\author{\small Linlin Ye$^1$, Zhaoqi Wu$^1$\thanks{Corresponding author. E-mail: wuzhaoqi\_conquer@163.com},
Shao-Ming Fei$^{2,3}$\\
{\small\it  1. Department of Mathematics, Nanchang University,
Nanchang 330031, P R China}\\
{\small\it  2. School of Mathematical Sciences, Capital Normal
University, Beijing 100048, P R China}\\
{\small\it 3. Max Planck Institute for Mathematics in the Sciences,
04103 Leipzig, Germany}}

\maketitle

\noindent {\bf Abstract} {\small }\\
Shor's algorithm outperforms its classical counterpart in efficient
prime factorization. We explore the coherence and entanglement
dynamics of the evolved states within Shor's algorithm, showing that
the coherence in each step relies on the dimension of register or
the order, and discuss the relations between geometric coherence and
geometric entanglement. We investigate how unitary
operators induce variations in coherence and entanglement, and
analyze the variations of coherence and entanglement within the
entire algorithm, demonstrating that the overall effect of Shor's
algorithm tends to deplete coherence and produce entanglement. Our
research not only deepens the understanding of this algorithm but
also provides methodological references for studying resource
dynamics in other quantum algorithms.

\noindent {\bf Keywords}: Shor's algorithm; Tsallis relative
$\alpha$ entropy of coherence; $l_{1,p}$ norm of coherence;
Geometric coherence; Geometric entanglement
\vskip0.2in

\noindent {\bf 1 Introduction}\\\hspace*{\fill}\\
Quantum coherence, regarded as a crucial resource, has become a
fundamental concept in the field of quantum information processing.
It originates from the field of optics and is grounded in the
principle of quantum superposition, which is utilized to elucidate
the phenomenon of light interference\cite{SEC,GRJ,SMO}. A
foundational framework for quantifying coherence as a resource has
been introduced in\cite{TB}, with an accessible and practical
analogous formulation provided in\cite{YXD}. Based on these
frameworks, coherence measures have been presented from the
perspective of information\cite{RSP,SLH,YCSQ,ZXN,XJC}, such as
geometric coherence\cite{SAS}, Tsallis relative $\alpha$
entropy of coherence\cite{ZHYC} and $l_{1,p}$ norm of
coherence\cite{JYL}. Quantification of coherence in
infinite-dimensional systems has also been considered\cite{JXP,YRZ}.
On the other hand, coherence manipulation has also attracted much
attention and fruitful results have been
obtained\cite{Dushuanping1,Dushuanping2,Dushuanping3,Dushuanping4}.
Coherence has found extensive utility across various domains,
including biological systems\cite{HS,LS}, nanoscale
physics\cite{KOL}, quantum phase transitions\cite{JJ} and
thermodynamic systems\cite{LM,MLK,VNG,MHJ,PK}.

The fidelity-based geometric coherence has been
established as a full convex coherence monotone\cite{SAS}, offering
the advantage of numerical computation through semidefinite
programming for finite-dimensional mixed states\cite{ZZD}. Notably,
this measure possesses a closed-form expression for single-qubit
states, enhancing its practical utility. The $l_{q,p}$
norm\cite{KAL}, which depends solely on the spatial structure of
matrices, has been demonstrated to induce a well-defined coherence
measure when $q=1$ and $p\in[1,2]$\cite{JYL}, providing a simple
closed form incorporating previous norm-induced coherence measures.
Regarding entropy-based approaches, Tsallis relative $\alpha$
entropy has been extensively explored as a measure for assessing the
purity of a state\cite{AS1,AS2}. Coherence based on it was initially
proposed in \cite{RAEQ}, and later rectified by Zhao and Yu
\cite{ZHYC} to be a rigorous coherence measure. The Tsallis relative
$\alpha$ entropy of coherence and the $l_{1}$ norm of coherence
yield identical ordering for single-qubit pure states \cite{FGZ},
suggesting deeper mathematical relationships between seemingly
distinct coherence quantifiers.

Entanglement, a unique bond among quantum particles, has been the
subject of extensive research over the years. Initially, it was
primarily understood as a qualitative aspect of quantum theory,
closely related to quantum nonlocality and Bell's
inequality\cite{BLER,BJSO}. However, it has also proven to be
instrumental in certain contexts like quantum cryptography
\cite{BCHW, EAKQ} and is considered as a pivotal resource in quantum
computation and information \cite{MA, BDC}. The geometric
entanglement was initially proposed for bipartite pure
states\cite{SADE}, and later extended to the multipartite
scenario\cite{BHLN} using projectors of varying ranks, which was
proven to be an appropriate entanglement quantifier, particularly
when dealing with multipartite systems \cite{WTC}. Furthermore, it
is a compelling entanglement measure due to its links with various
other measures\cite{CDC} and the fact that it can be efficiently
estimated using quantitative entanglement witnesses, which are
suitable for experimental validation\cite{GOR}.

Quantum algorithms can enhance the efficiency of problem-solving on
quantum computers\cite{MA,ZhouNR}. Many quantum algorithms, such as
Deutsch-Jozsa algorithm\cite{DDJ,CDKK}, Simon's
algorithm\cite{SDR,TMS,GSP}, Grover's search algorithm\cite{GLK} ,
Bernstein-Vazirani algorithm\cite{DJSM} and HHL algorithm \cite{HAW}
have been proposed. Shor's algorithm\cite{PWSP} enables the rapid
factorization of large numbers, allowing for the decryption of
RSA-encrypted communications. The factoring problem can be reduced
to a problem of finding the order $r$, the period of $x^{a}
\mathrm{mod} N$, where $N$ is the number to be factored, and $x$ is
an integer coprime to $N$. Monz\cite{MTE} utilized a miniature
quantum computer comprising five trapped calcium ions to execute a
scalable version of Shor's algorithm. This approach was more
efficient than previous implementations and held promise for the
design of powerful quantum computers that optimized the use of
available resources.

The dynamics of entanglement and coherence within Grover's algorithm
has been the subject of extensive
investigation\cite{MinghuaPan2017QIP,MinghuaPan2019TCS,RBD,SHL,PMQ,MPH,YLWZ1,Sun2024,FCC}.
Coherence dynamics in Deutsch-Jozsa algorithm, Simon's quantum
algorithm and HHL quantum algorithm have also been investigated in
recent years\cite{FCC,HMC,LYC,YLWZ2,YLWZ3}. Ahnefeld examined a
sequential version of Shor's algorithm that maintains a consistent
overall structure, and utilize channel coherence to derive a lower
and an upper bound on the success probability of the
protocol.\cite{AFTTE}. Naseri et al.\cite{NMKT} investigated the
roles of coherence and entanglement as quantum resources in the
Bernstein-Vazirani algorithm. In cases where there is no
entanglement in the initial and final states of the algorithm, the
performance of the algorithm is directly related to the coherence of
the initial state. A significant amount of geometric entanglement
can prevent the algorithm from achieving optimal performance.


Coherence and entanglement are regarded as the key factors
underlying the superior efficiency of quantum algorithms over
classical methods in specific computational contexts. By examining
the dynamic evolution of quantum resource throughout the algorithms,
we can elucidate how quantum superposition and interference
facilitate computational speedup, thereby addressing the fundamental
question of how quantum resources translate into computational
advantages. This analysis not only deepens our theoretical
understanding of quantum computation but also provides practical
insights for optimizing resource allocation in future quantum
algorithmic designs. Our study on quantum coherence and entanglement
dynamics in Shor's algorithm reveals a fundamental relationship
between these resources and success probability. The results
demonstrate that coherence consumption necessarily precedes
probability enhancement, while the algorithm effectively transforms
coherence into entanglement critical for quantum speedup.

The remainder of this paper is structured as follows. In Section 2,
we recall Shor's algorithm and several coherence and
entanglement measures. In Section 3, we investigate the coherence
and entanglement dynamics in Shor's algorithm. We explore
the variations of coherence and
entanglement in Shor's algorithm in Section 4. In Section 5,
we choose a specific example to illustrate the coherence/
entanglement dynamics. Finally, we summarize our results in Section 6.

\vskip0.1in

\noindent {\bf 2 Shor's algorithm and coherence/entanglement quantifiers}\\\hspace*{\fill}\\
In this section, we recall Shor's algorithm and the
quantifiers of coherence and entanglement we use in this paper.
\vskip0.1in

\noindent {\bf 2.1 Shor's algorithm}\\\hspace*{\fill}\\
The problem of prime factorization is formulated as follows. Given
an odd composite positive integer $N$, the task is to determine its
prime factors. It is well known that factoring a composite number
$N$ reduces to the order-finding problem. Let $N$ be an integer and
$x$ be an integer such that $x<N$ and $x$ is coprime to $N$. The
objective is to find the order $r$, which is the smallest positive
integer satisfying $x^r=1\pmod{N}$. To deal with this problem the
qubit systems comprise a total of $n$ qubits, which are divided into
two registers $A$ and $B$. Register $A$ consists of $t$ qubits with dimension $Q=2^{t}$, while register $B$ contains $L$ qubits, where $t = 2L + 1 + \lceil\log(2 + \frac{1}{2\epsilon})\rceil$. Initialize the combined system $AB$ in the
state $|0\rangle_{A}^{\otimes t}|1\rangle_{B}$.
The main steps of the order-finding algorithm can be summarized as follows\cite{PWSP,Qiu2023}:\\
(i) Apply Hadamard gates $H^{\otimes
t}=\frac{1}{\sqrt{Q}}\sum_{x,y}(-1)^{xy}|y\rangle\langle x|$ to the
qubits in register $A$. Then one gets
\begin{equation}\label{eq1}
|\psi_{1}\rangle=\frac{1}{\sqrt{Q}}\sum_{j=0}^{Q-1}|j\rangle|1\rangle.
\end{equation}
(ii) Apply the unitary transformation $U=\sum_{n=0}^{Q-1}|n\rangle\langle n|_{A}\otimes U_{B}^{n}$ on $|\psi_{1}\rangle$, where $U|j\rangle|y\rangle = |j\rangle|x^j y \bmod N\rangle$ and $U_{B}|y\rangle_{B}=|xy \bmod N\rangle_{B}$. Then one obtains the state
\begin{equation}\label{eq2}
|\psi_{2}\rangle=\frac{1}{\sqrt{Q}}\sum_{j=0}^{Q-1}|j\rangle|x^{j} \mathrm{mod} N\rangle.
\end{equation}
Let the eigenvectors of $U_{B}$ be
\begin{equation}\label{eq3}
|u_{s}\rangle=\frac{1}{\sqrt{r}}\sum_{a=0}^{r-1}\mathrm{e}^{-\frac{2\pi
\mathrm{i}as}{r}}|x^{a} \mathrm{mod} N\rangle.
\end{equation}
In particular, one has $\frac{1}{\sqrt{r}}\sum_{s=0}^{r-1}|u_{s}\rangle=|1\rangle$.
The eigenvalues corresponding to $|u_{s}\rangle$ are $\mathrm{e}^{\frac{2\pi \mathrm{i}s}{r}}$.
Then we have
\begin{equation}\label{eq4}
|\psi_{2}\rangle\approx U\frac{1}{\sqrt{Q}} \sum_{j=0}^{Q-1} |j\rangle|1\rangle = \frac{1}{\sqrt{Q}} \sum_{j=0}^{Q-1} |j\rangle U_{B}^{j}\left(\frac{1}{\sqrt{r}} \sum_{s=0}^{r-1} |u_s\rangle\right) = \frac{1}{\sqrt{Qr}} \sum_{s=0}^{r-1} \sum_{j=0}^{Q-1}\mathrm{e}^{2\pi \mathrm{i}j \frac{s}{r}} |j\rangle |u_s\rangle,
\end{equation}
where $0\leq s\leq r-1$.\\
(iii) Applying the inverse Fourier transform $F^\dagger$ to register $A$, one has
\begin{align}\label{eq5}
|\psi_{3}\rangle
&=(F^\dagger\otimes I)\frac{1}{\sqrt{Qr}} \sum_{s=0}^{r-1} \sum_{j=0}^{Q-1}\mathrm{e}^{2\pi \mathrm{i}j \frac{s}{r}} |j\rangle |u_s\rangle=\frac{1}{\sqrt{r}} \sum_{s=0}^{r-1} F^\dagger\frac{1}{\sqrt{Q}}\sum_{j=0}^{Q-1}\mathrm{e}^{2\pi \mathrm{i}j \frac{s}{r}} |j\rangle (I|u_s\rangle)\notag\\
&=\frac{1}{\sqrt{r}}\sum_{s=0}^{r-1}|\psi_{s}\rangle|u_{s}\rangle
\end{align}
where $|\psi_{s}\rangle = F^\dagger\frac{1}{\sqrt{Q}} \sum_{j=0}^{Q-1} \mathrm{e}^{2\pi \mathrm{i} j \frac{s}{r}} |j\rangle=\frac{1}{Q} \sum_{j=0}^{Q-1}\sum_{k=0}^{Q-1} \mathrm{e}^{2\pi \mathrm{i} j (\frac{s}{r}-\frac{k}{Q})} |k\rangle$.
We measure the first register to obtain an approximation $\frac{k}{Q} \approx \frac{s}{r} (s \in \{0, 1, \ldots, r-1\})$.
$|k\rangle$ is a $t$-bit string that is an estimation of $\frac{s}{r}$ for some $s$.
More specifically, the probability of obtaining measurement outcome $k$ from the first register in step (iii) is given by
\begin{equation}\label{eq6}
p_k = \frac{1}{r} \sum_{s=0}^{r-1} \left| \frac{1}{Q}\sum_{j=0}^{Q-1}
\mathrm{e}^{2\pi \mathrm{i} j (\frac{s}{r}-\frac{k}{Q})} \right|^2.
\end{equation}
Intuitively, the order-finding algorithm randomly selects a value $s \in \{0, 1, \ldots, r-1\}$ and then returns an approximation to $\frac{s}{r}$ in the form $\frac{k}{Q}$. Finally, by applying the continued fraction algorithm, the order $r$ can be found.

\vskip0.1in
\noindent {\bf 2.2 Measures of coherence and entanglement}\\\hspace*{\fill}\\
{\bf Definition 1} The Tsallis relative $\alpha$ entropy is defined
as\cite{AS1,AS2}
\begin{equation}\label{eq7}
D_{\alpha}(\rho\|\sigma)=\frac{1}{\alpha-1}\left(f_{\alpha}(\rho,\sigma)
-1\right),
\end{equation}
where $\alpha\in(0,1)\cup(1,\infty)$ and $f_{\alpha}(\rho,\sigma)=\mathrm{Tr}(\rho^{\alpha}\sigma^{1-\alpha})$.
Note that when $\alpha\rightarrow1$, $D_{\alpha}(\rho\|\sigma)$ reduces to
$S'(\rho\parallel\sigma)=\ln2\cdot S(\rho\parallel\sigma)$, with
$S(\rho\parallel\sigma)= \mathrm{Tr}(\rho\log\rho)-
\mathrm{Tr}(\rho\log\sigma)$ representing the standard relative entropy. Here,
the logarithm `log' is taken to be base 2.

\noindent {\bf Definition 2} Let $\{|i\rangle\}_{i=1}^d$ be an
orthonormal basis in a $d$ dimensional Hilbert space $H$. For
$\alpha \in (0,1) \cup (1,2]$, the Tsallis relative $\alpha$-entropy
of coherence has been defined by \cite{ZHYC}
\begin{equation}\label{eq8}
C_{\alpha}(\rho)=\mathop{\mathrm{min}}\limits_{\sigma\in
\mathcal{I}}\frac{1}{\alpha-1}\left(f_{\alpha}^{\frac{1}{\alpha}}(\rho,\sigma)
-1\right)=\frac{1}{\alpha-1}\left[\sum_{i=1}^{d}\langle
i|\rho^{\alpha}|i\rangle^{\frac{1}{\alpha}}-1\right],
\end{equation}
and was proven to be a family of bona-fide coherence measures, where
$\mathcal{I}$ denotes the set of incoherent states, which are
diagonal matrices in the given basis. Note that when
$\alpha\rightarrow1$, $C_{\alpha}(\rho)$ reduces to $\ln2\cdot
C_{r}(\rho)$, where $C_{r}(\rho)=\mathrm{Tr}(\rho\log\rho)-
\mathrm{Tr}(\rho_{\mathrm{diag}}\log\rho_{\mathrm{diag}})$ is the
relative entropy of coherence \cite{TB}, with $\rho_{\mathrm{diag}}$
denotes the diagonal part of the state $\rho$. Also,
$C_{\alpha}(\rho)$ reduces to $2C_{s}(\rho)$ when
$\alpha=\frac{1}{2}$, where $C_{s}(\rho)=1-\sum_{j=1}^{d}\langle
j|\sqrt{\rho}|j\rangle^{2}$ is the skew information of coherence
\cite{YCSQ}.

\noindent {\bf Definition 3}
The $l_{q,p}$ norm of a matrix $A \in M_n$ is defined as the $l_q$ norm of the vector obtained by applying the $l_p$ norm to the columns of $A$, given by \cite{JYL}
\begin{equation}\label{eq9}
l_{q,p}(A)=\left(\sum_{j=1}^{n}l_{p}(A_{j})^{q}\right)^{\frac{1}{q}},
\end{equation}
where $1\leq p$, $q\leq\infty$, $A_{j}$ denotes the columns of $A$, and
$l_{p}(A_{j})=\left(\sum_{i=1}^{n}|A_{i,j}|^{p}\right)^{\frac{1}{p}}$,
with $A_{i,j}$ being the entry of the $i$th row and $j$th column of matrix $A$.

Note that $l_{p,p}$ norm is actually the $l_p$ norm. The coherence
based on $l_{q,p}$ norm is a well-defined coherence measure if and
only if $q=1$ and $p\in[1,2]$.

\noindent {\bf Definition 4} The $l_{1,p}$ norm of coherence
$C_{1,p}$ of a density operator $\rho$ for $p\in[1,2]$ is defined
by\cite{JYL}
\begin{align}\label{eq10}
C_{1,p}(\rho)\notag
=&\mathop{\min}\limits_{\sigma\in\mathcal{I}}l_{1,p}(\rho-\sigma)
=l_{1,p}(\rho-\rho_\mathrm{diag})\\
=&\sum_{j=1}^{n}\left(\sum_{i=1}^{n}|(\rho-\rho_\mathrm{diag})_{i,j}|^{p}\right)^{\frac{1}{p}},
\end{align}
where $(\rho-\rho_\mathrm{diag})_{i,j}$ is the $(i,j)$-th entry of
$\rho-\rho_\mathrm{diag}$.

The coherence $C_{1,p}$ introduces a class of coherence measures
that hold potential utility and expands the methodological framework
for analyzing quantum coherence in multipartite systems. It is
worthwhile to note that when $p=q=1$, $C_{1,p}(\rho)$ reduces to the
$l_{1}$ norm of coherence $C_{l_{1}}(\rho)=\sum_{i\neq
j}|\rho_{ij}|$\cite{TB}.

The fidelity between two quantum states $\rho$ and $\sigma$ is
defined by
$F(\rho,\sigma)=\mathrm{Tr}\sqrt{\rho^{\frac{1}{2}}\sigma\rho^{\frac{1}{2}}}$\cite{MA}.
Particularly, when $\rho=|\psi\rangle \langle\psi|$ is a pure state,
we have
$F(|\psi\rangle,\sigma)=\mathrm{Tr}\sqrt{\langle\psi|\sigma|\psi\rangle}$.

\noindent {\bf Definition 5} The geometric coherence is
defined by\cite{SAS}
\begin{equation}\label{eq11}
C_{g}(\rho)=1-\left[\mathop{\mathrm{max}}\limits_{\delta\in \mathcal{I}}F(\rho,\delta)\right]^{2}.
\end{equation}
For a pure state  $|\psi\rangle=\sum_{i}\lambda_{i}|i\rangle$, its geometric coherence is given by
\begin{equation}\label{eq12}
C_{g}(\rho)=1-\mathop{\mathrm{max}}\limits_{i}\{|\lambda_{i}|^{2}\},
\end{equation}
where $|\lambda_{i}|^{2}$ are the diagonal elements of $|\psi\rangle\langle\psi|$.

\noindent {\bf Definition 6} The geometric entanglement
for a state $\rho=|\psi\rangle\langle\psi|$ is defined as the
minimum distance to any separable state $|\phi\rangle$, which
corresponds to the maximum overlap between $|\psi\rangle$ and
$|\phi\rangle$\cite{SADE},
\begin{equation}\label{eq13}
E_{g}(\rho)=1-\mathop{\max}\limits_{|\phi\rangle}|\langle\psi|\phi\rangle|^{2},
\end{equation}
where the maximum is taken over all separable states
$|\phi\rangle=\bigotimes_{s=1}^{n}|\phi_{s}\rangle$, with
$|\phi_{s}\rangle$ denoting the single-qubit pure states.

\vskip0.1in

\noindent {\bf 3 Coherence and entanglement dynamics in Shor's algorithm}\\\hspace*{\fill}\\
In this section, we examine the dynamics of coherence and entanglement in the state resulting from the application of each basic operator within Shor's algorithm.

Let $\rho_{1}=|\psi_{1}\rangle\langle\psi_{1}|$ be the density operator of the state $|\psi_{1}\rangle$. By employing Eqs. (\ref{eq1}), (\ref{eq8}), (\ref{eq10}) (\ref{eq12}) and (\ref{eq13}) we have\\\hspace*{\fill}\\
{\bf Theorem 1} The $l_{1,p}$ norm of coherence, the Tsallis relative $\alpha$ entropy of coherence, the geometric coherence and the geometric entanglement of the state $\rho_{1}$ are given by
\begin{equation}\label{eq14}
C_{1,p}(\rho_{1})=\left(Q-1\right)^{\frac{1}{p}},
\end{equation}
\begin{equation}\label{eq15}
C_{\alpha}(\rho_{1})=\frac{1}{\alpha-1}\left(Q^{1-\frac{1}{\alpha}}-1\right),
\end{equation}
\begin{equation}\label{eq16}
C_{g}(\rho_{1})=1-\frac{1}{Q}
\end{equation}
and
\begin{equation}\label{eq17}
E_{g}(\rho_1)=0,
\end{equation}
respectively.

\noindent {\bf Remark 1} According to Theorem 1, the $l_{1,p}$ norm
of coherence, the Tsallis relative $\alpha$ entropy of coherence,
and the geometric coherence of $\rho_{1}$ are all dependent on the
dimension of register $Q$, while the entire system represented by
$\rho_1$ does not manifest any entanglement. Also, note that
\begin{equation*}
C_{g}(\rho_{1})+E_{g}(\rho_1)<1.
\end{equation*}

The optimization of the geometric entanglement is achievable for a
subset of symmetric separable states $|\eta\rangle^{\otimes n}$,
where
$|\eta\rangle=\cos\frac{\alpha}{2}|0\rangle+\mathrm{e}^{\mathrm{i}\beta}\sin\frac{\alpha}{2}|1\rangle$,
with $\alpha\in [0, \pi]$, $\beta\in[0, 2\pi]$ and $\mathrm{i}$ the
imaginary unit. Since the coefficients of $|\psi_{2}\rangle$ are all
positive, one sets $\beta=0$. To simplify the
analysis, we consider the case where $Q = rm$ with $m$ being a
positive integer, which implies that
$\left\lfloor\frac{Q-1}{r}\right\rfloor = m-1$. Expressing the index
as $j = a + br$, the state $|\psi_{2}\rangle$ can be approximately
expressed as
$\frac{1}{\sqrt{Q}}\sum_{a=0}^{r-1}\sum_{b=0}^{m-1}|a+br\rangle|x^{a}
\mathrm{mod} N\rangle$, where $0\leq a\leq r-1$ and $0\leq b\leq
m-1$. Letting
$|S_{a}\rangle=\sum_{b=0}^{m-1}|a+br\rangle|x^{a}\mathrm{mod}
N\rangle$, we can then rewrite $|\psi_{2}\rangle$ as
\begin{equation*}
\sqrt{\frac{1}{Q}}\sum_{a=0}^{r-1}|S_{a}\rangle.
\end{equation*}
Recall that the Hamming weight of a basis state $|x\rangle$ is the
number of 1's in the binary string $x\in\{0,1\}^{n}$, and we use
$|x|$ to denote the Hamming weight of
$|x\rangle$\cite{MinghuaPan2019TCS}.

Let $\rho_{2}=|\psi_{2}\rangle\langle\psi_{2}|$ be the density operator of the state $|\psi_{2}\rangle$, and the following result then holds.\\
{\bf Theorem 2} The Tsallis relative $\alpha$ entropy of coherence, the $l_{1,p}$ norm of coherence and the geometric coherence of the state $\rho_2$ are the
same as the ones of $\rho_1$. The geometric entanglement of the state $\rho_2$ is given by
\begin{equation}\label{eq18}
E_{g}(\rho_2)=1-\frac{1}{Q}\left(\sum_{a=0}^{r-1}\sum_{b=0}^{m-1}\left(\frac{n-n_{a,b}}{n}\right)
^{\frac{n-n_{a,b}}{2}}
\left(\frac{n_{a,b}}{n}\right)^{\frac{n_{a,b}}{2}}\right)^{2},
\end{equation}
where $n_{a,b}$ is the Hamming weight of $|a+br\rangle|x^{a}\mathrm{mod} N\rangle(a=0,1,\cdots,r-1$ and $b=0,1,\cdots,m-1)$.\\\hspace*{\fill}\\
$\it{Proof}$. Direct calculations show that
\begin{equation*}
C_{1,p}(\rho_{1})=C_{1,p}(\rho_{2}),~~~ C_{\alpha}(\rho_{1})=C_{\alpha}(\rho_{2})~~~
\mathrm{and}~~~
C_{g}(\rho_{2})=C_{g}(\rho_{1}).
\end{equation*}
The symmetric $n$-separable state can be expressed as
\begin{equation}
|\phi\rangle=|\eta\rangle^{\otimes
n}=\sum_{x\in\{0,1\}^{n}}\cos^{n-|x|}\frac{\alpha}{2}\sin^{|x|}\frac{\alpha}{2}|x\rangle,
\nonumber
\end{equation}
where $|x|$ represents the Hamming weight of $|x\rangle$. According
to \cite{MinghuaPan2017QIP}, the overlap between the state
$|\psi_{2}\rangle$ and the separable state $|\phi\rangle$ is
\begin{equation*}
\langle\psi_{2}|\phi\rangle=\sqrt{\frac{1}{Q}}\sum_{a=0}^{r-1}
\left(\sum_{b=0}^{m-1}\cos^{n-n_{a,b}}\frac{\alpha}{2}\sin^{n_{a,b}}\frac{\alpha}{2}\right),
\end{equation*}
where $n_{a,b}$ is the Hamming weight of $|a+br\rangle|x^{a}\mathrm{mod} N\rangle$. Then the maximum overlap between the state $|\psi_{2}\rangle$ and the
separable state $|\phi\rangle$ is
\begin{equation*}
\max\limits_{|\phi\rangle}|\langle\psi_{2}|\phi\rangle|^{2}=\max\limits_{\alpha}
\left|\frac{1}{\sqrt{Q}}\sum_{a=0}^{r-1}\sum_{b=0}^{m-1}
\cos^{n-n_{a,b}}\frac{\alpha}{2}\sin^{n_{a,b}}\frac{\alpha}{2}\right|^{2}.
\end{equation*}
Denote
$A(\alpha)=\cos^{n-n_{a,b}}\frac{\alpha}{2}\sin^{n_{a,b}}\frac{\alpha}{2}$.
The maximum of $A(\alpha)$ is
\begin{equation*}\label{eq}
A(\alpha)_{\max}=\left(\frac{n-n_{a,b}}{n}\right)^{\frac{n-n_{a,b}}{2}}
\left(\frac{n_{a,b}}{n}\right)^{\frac{n_{a,b}}{2}}.
\end{equation*}
Therefore, we get
\begin{equation*}
\max\limits_{|\phi\rangle}|\langle\psi_{2}|\phi\rangle|^{2}=\frac{1}{Q}
\left(\sum_{a=0}^{r-1}\sum_{b=0}^{m-1}
\left(\frac{n-n_{a,b}}{n}\right)^{\frac{n-n_{a,b}}{2}}
\left(\frac{n_{a,b}}{n}\right)^{\frac{n_{a,b}}{2}}\right)^{2},
\end{equation*}
which is attained at $\alpha=2\arccos\sqrt{\frac{n-n_{a,b}}{n}}$. From
Eq. (\ref{eq13}), we then have
\begin{equation*}
E_{g}(\rho_2)=1-\frac{1}{Q}\left(\sum_{a=0}^{r-1}\sum_{b=0}^{m-1}\left(\frac{n-n_{a,b}}{n}\right)
^{\frac{n-n_{a,b}}{2}}
\left(\frac{n_{a,b}}{n}\right)^{\frac{n_{a,b}}{2}}\right)^{2}.
\end{equation*}
$\hfill\qedsymbol$ \\\hspace*{\fill}\\
{\bf Remark 2} It can be seen from Theorem 2 that the
$E_{g}(\rho_2)$ depends on the dimension of register $A$, the total
number of qubits in the system, the Hamming weight of
$|a+br\rangle|x^{a}\mathrm{mod} N\rangle$ and the order $r$, while
both $E_g(\rho_2)$ and $C_g(\rho_2)$ are less than 1 and dependent
on $Q$. Moreover, we have
\begin{equation*}
C_g(\rho_2)+\frac{1}{\gamma(n,n_{a,b})}(1-E_{g}(\rho_2))=1,
\end{equation*}
where $\gamma(n,n_{a,b})=\left(\sum_{a=0}^{r-1}\sum_{b=0}^{m-1}\left(\frac{n-n_{a,b}}{n}\right)
^{\frac{n-n_{a,b}}{2}}
\left(\frac{n_{a,b}}{n}\right)^{\frac{n_{a,b}}{2}}\right)^{2}$ and $0<\gamma(n,n_{a,b})<Q$.
It is easy to see that $E_{g}(\rho_2)$ becomes larger if
$C_g(\rho_2)$ is larger, while $C_g(\rho_2)>E_{g}(\rho_2)$ when
$\gamma(n,n_{a,b})>1$, $C_g(\rho_2)<E_{g}(\rho_2)$ when
$\gamma(n,n_{a,b})<1$ and $C_g(\rho_2)=E_{g}(\rho_2)$ when
$\gamma(n,n_{a,b})=1$.


For a given $s \in \{0, 1, \ldots, r-1\}$, to
simplify the analysis, we consider the ideal case in which there
exists an integer $k_s$ such that $\frac{k_s}{Q} = \frac{s}{r}$
holds, i.e., $k_s = \frac{sQ}{r}=sm$. Then the expression for
$|\psi_s\rangle$ can be simplified to $|\psi_s\rangle =
F^\dagger\frac{1}{\sqrt{Q}} \sum_{j=0}^{Q-1} \mathrm{e}^{2\pi \mathrm{i} j
\frac{k_s}{Q}} |j\rangle = |k_s\rangle$, and thus $|\psi_{3}\rangle$
can be rewritten as
\begin{equation}\label{eq19}
|\psi_{3}\rangle
=\frac{1}{\sqrt{r}}\sum_{s=0}^{r-1}|\psi_s\rangle|u_{s}\rangle
=\frac{1}{\sqrt{r}}\sum_{s=0}^{r-1}|k_{s}\rangle|u_{s}\rangle
=\frac{1}{r}\sum_{s,a=0}^{r-1}\mathrm{e}^{-\frac{2\pi
\mathrm{i}as}{r}}|k_{s}\rangle|x^{a} \mathrm{mod} N\rangle.
\end{equation}

Let $\rho_{3}$ be the density operator of the state $|\psi_{3}\rangle$. We have by direct calculation the following theorem.\\\hspace*{\fill}\\
{\bf Theorem 3} The $l_{1,p}$ norm of coherence, the Tsallis relative $\alpha$ entropy of coherence, the geometric coherence and the geometric entanglement of the state $\rho_{3}$ are given by
\begin{equation}\label{eq20}
C_{1,p}(\rho_{3})=\left(r^{2}-1\right)^{\frac{1}{p}},
\end{equation}
\begin{equation}\label{eq21}
C_{\alpha}(\rho_{3})=\frac{1}{\alpha-1}\left(r^{2(1-\frac{1}{\alpha})}-1\right),
\end{equation}
\begin{equation}\label{eq22}
C_g(\rho_{3})=1-\frac{1}{r^{2}}
\end{equation}
and
\begin{equation}\label{eq23}
E_g(\rho_3)=1-\frac{1}{r^{2}}\left(\sum_{a,s=0}^{r-1}\mathrm{e}^{-\frac{2\pi \mathrm{i}sa}{r}}\left(\frac{n-m_{a,s}}{n}\right)
^{\frac{n-m_{a,s}}{2}}
\left(\frac{m_{a,s}}{n}\right)^{\frac{m_{a,s}}{2}}\right)^{2},
\end{equation}
respectively, where $m_{a,s}$ is the Hamming weight of $|k_{s}\rangle|x^{a}\mathrm{mod} N\rangle$ $(a,s=0,1,\cdots,r-1)$.\\\hspace*{\fill}\\
$\it{Proof}$. By substituting Eq.(\ref{eq19}) into Eqs. (\ref{eq8}), (\ref{eq10}) and (\ref{eq12}),
we obtain Eqs. (\ref{eq20}), (\ref{eq21}) and (\ref{eq22}).
Rewrite $|\psi_{3}\rangle=\frac{1}{r}\sum_{a,s=0}^{r-1}\mathrm{e}^{-\frac{2\pi \mathrm{i}sa}{r}}|k_{s}\rangle|x^{a}\mathrm{mod} N\rangle$ as
\begin{equation*}
\frac{1}{r}\sum_{a,s=0}^{r-1}\mathrm{e}^{-\frac{2\pi \mathrm{i}sa}{r}}|W_{a,s}\rangle,
\end{equation*}
where $|W_{a,s}\rangle=|k_{s}\rangle|x^{a}\mathrm{mod} N\rangle$.
According to \cite{MinghuaPan2017QIP}, the overlap between the
separable state $|\phi\rangle$ and the state $|\psi_{3}\rangle$ is
\begin{equation*}
\langle\psi_{3}|\phi\rangle=\frac{1}{r}\sum_{a,s=0}^{r-1}\mathrm{e}^{-\frac{2\pi
\mathrm{i}sa}{r}} \cos^{n-m_{a,s}}\frac{\alpha}{2}\sin^{m_{a,s}}\frac{\alpha}{2},
\end{equation*}
where $m_{a,s}$ is the Hamming weight of $|k_{s}\rangle|x^{a}\mathrm{mod} N\rangle$. Denote
$B(\alpha)=\cos^{n-m_{a,s}}\frac{\alpha}{2}$ $\sin^{m_{a,s}}\frac{\alpha}{2}$.
It is verified that when $\alpha=2\arccos\sqrt{\frac{n-m_{a,s}}{n}}$,
the maximum value of $B(\alpha)$ is
\begin{equation*}
B(\alpha)_{\max}=\left(\frac{n-m_{a,s}}{n}\right)^{\frac{n-m_{a,s}}{2}}
\left(\frac{m_{a,s}}{n}\right)^{\frac{m_{a,s}}{2}}.
\end{equation*}
Then the maximum of the overlap between the separable state $|\phi\rangle$
and the state $|\psi_{3}\rangle$ is
\begin{align*}
\max\limits_{|\phi\rangle}|\langle\psi_{3}|\phi\rangle|^{2}\notag
=&\max\limits_{\alpha}\left|\frac{1}{r}\sum_{a,s=0}^{r-1}\mathrm{e}^{-\frac{2\pi \mathrm{i}sa}{r}}\cos^{n-m_{a,s}}\frac{\alpha}{2}\sin^{m_{a,s}}\frac{\alpha}{2}\right|^{2}\\
=&\frac{1}{r^{2}}\left(\sum_{a,s=0}^{r-1}\mathrm{e}^{-\frac{2\pi \mathrm{i}sa}{r}}\left(\frac{n-m_{a,s}}{n}\right)^{\frac{n-m_{a,s}}{2}}
\left(\frac{m_{a,s}}{n}\right)^{\frac{m_{a,s}}{2}}\right)^{2}.
\end{align*}
Hence, according to Eq. (\ref{eq13}) we have
\begin{equation*}
E_{g}(\rho_3)=1-\frac{1}{r^{2}}\left(\sum_{a,s=0}^{r-1}\mathrm{e}^{-\frac{2\pi \mathrm{i}sa}{r}}\left(\frac{n-m_{a,s}}{n}\right)
^{\frac{n-m_{a,s}}{2}}
\left(\frac{m_{a,s}}{n}\right)^{\frac{m_{a,s}}{2}}\right)^{2}.
\end{equation*}
$\hfill\qedsymbol$ \\\hspace*{\fill}\\
{\bf Remark 3} Theorem 3 indicates that the $l_{1,p}$ norm of
coherence, the Tsallis relative $\alpha$ entropy of coherence and
the geometric coherence of the state $\rho_3$ are all determined by
the order $r$, while the geometric entanglement of the state
$\rho_3$ depends on the total number of qubits in the system, the
Hamming weight of $|k_{s}\rangle|x^{a}\mathrm{mod} N\rangle$ and the
order $r$. Note that both $E_g(\rho_3)$ and $C_g(\rho_3)$ are less
than 1 and depend on $r$. Besides, it holds that
\begin{equation*}
C_g(\rho_3)+\frac{1}{\gamma(n,m_{a,s})}(1-E_{g}(\rho_3))=1,
\end{equation*}
where $\gamma(n,m_{a,s})=\left(\sum_{a,s=0}^{r-1}\mathrm{e}^{-\frac{2\pi
\mathrm{i}sa}{r}}\left(\frac{n-m_{a,s}}{n}\right) ^{\frac{n-m_{a,s}}{2}}
\left(\frac{m_{a,s}}{n}\right)^{\frac{m_{a,s}}{2}}\right)^{2}$ and
$0<\gamma(n,m_{a,s})<r^{2}$. Similarly, $E_{g}(\rho_3)$ becomes
larger if $C_g(\rho_3)$ is larger, while $C_g(\rho_3)>E_{g}(\rho_3)$
when $\gamma(n,m_{a,s})>1$, $C_g(\rho_3)<E_{g}(\rho_3)$ when
$\gamma(n,m_{a,s})<1$ and $C_g(\rho_3)=E_{g}(\rho_3)$ when
$\gamma(n,m_{a,s})=1$.


\vskip0.1in
\noindent {\bf 4 Variations of coherence and entanglement in Shor's algorithm }\\\hspace*{\fill}\\
In this section, we investigate how unitary operators induce variations in coherence and entanglement, and analyze the variations of coherence
and entanglement within the entire algorithm.

Following \cite{MinghuaPan2019TCS} and \cite{PMQ}, we define the
variation of coherence and entanglement induced by a unitary
operator $V$ on a state $|\psi\rangle$ to be
\begin{equation}\label{eq24}
\Delta C(V,\rho)\equiv C(V\rho V^\dag)-C(\rho),
\end{equation}
and
\begin{equation}\label{eq25}
\Delta E(V,\rho)\equiv E(V\rho V^\dag)-E(\rho),
\end{equation} respectively, and the variation of coherence and entanglement in Shor's
quantum algorithm to be
\begin{equation}\label{eq26}
\Delta C(\rho)\equiv C(\rho_3)-C(\rho_1),
\end{equation}
and
\begin{equation}\label{eq27}
\Delta E(\rho)\equiv E(\rho_3)-E(\rho_1),
\end{equation}
respectively.

Combining Eqs. (\ref{eq14})-(\ref{eq17}),
(\ref{eq20})-(\ref{eq23}) and Theorem 2,
we have\\
{\bf Theorem 4} The variations of coherence and entanglement induced
by each operator are
\begin{equation}\label{eq28}
\Delta C_{1,p}(U,\rho_1)=0,
\end{equation}
\begin{equation}\label{eq29}
\Delta
C_{1,p}(F^{\dag},\rho_2)=\left(r^{2}-1\right)^{\frac{1}{p}}-\left(Q-1\right)^{\frac{1}{p}},
\end{equation}
\begin{equation}\label{eq30}
\Delta C_\alpha(U,\rho_1)=0,
\end{equation}
\begin{equation}\label{eq31}
\Delta
C_\alpha(F^{\dag},\rho_2)=\frac{1}{\alpha-1}\left[r^{2(1-\frac{1}{\alpha})}-Q^{1-\frac{1}{\alpha}}\right],
\end{equation}
\begin{equation}\label{eq32}
\Delta C_{g}(U,\rho_1)=0,
\end{equation}
\begin{equation}\label{eq33}
\Delta C_{g}(F^{\dag},\rho_2)=\frac{1}{Q}-\frac{1}{r^{2}},
\end{equation}
\begin{equation}\label{eq34}
\Delta E_g(U,\rho_1)
=1-\frac{1}{Q}\left(\sum_{a=0}^{r-1}\sum_{b=0}^{m-1}\left(\frac{n-n_{a,b}}{n}\right)
^{\frac{n-n_{a,b}}{2}}
\left(\frac{n_{a,b}}{n}\right)^{\frac{n_{a,b}}{2}}\right)^{2}
\end{equation}
and
\begin{align}\label{eq35}
\Delta E_g(F^{\dag},\rho_2)
&=\frac{1}{Q}\left(\sum_{a=0}^{r-1}\sum_{b=0}^{m-1}\left(\frac{n-n_{a,b}}{n}\right)
^{\frac{n-n_{a,b}}{2}}
\left(\frac{n_{a,b}}{n}\right)^{\frac{n_{a,b}}{2}}\right)^{2}\notag\\
&-\frac{1}{r^{2}}\left(\sum_{a,s=0}^{r-1}\mathrm{e}^{-\frac{2\pi \mathrm{i}sa}{r}}\left(\frac{n-m_{a,s}}{n}\right)
^{\frac{n-m_{a,s}}{2}}
\left(\frac{m_{a,s}}{n}\right)^{\frac{m_{a,s}}{2}}\right)^{2},
\end{align}
respectively.

In Shor's algorithm, $Q$ represents the dimension of register $A$ and is specifically chosen to satisfy the constraint $N^2 \leq Q < 2N^2$.
Combining this with the fact that $r \leq N$, we can derive the inequality $Q \geq r^2$. This relationship leads to $\Delta
C_{1,p}(F^{\dag},\rho_2)<0$, $\Delta
C_\alpha(F^{\dag},\rho_2)<0$, $\Delta C_{g}(F^{\dag},\rho_2)<0$ and $\Delta E_g(U,\rho_1)\geq0$.
Our analysis further reveals a distinct characterization of the operators' effects within the algorithm. Specifically, while the operator $U$ preserves the coherence of state $\rho_{1}$, it simultaneously generates entanglement. In contrast, the operator $F^\dag$ demonstrably consumes coherence throughout its application. Furthermore, the behavior of $F^{\dag}$ with respect to entanglement exhibits more nuanced properties and demonstrates condition-dependency, suggesting a complex interplay between different quantum resources during computational processes.

\vskip0.1in

Based on Eqs. (\ref{eq26}) and (\ref{eq27}) and Theorem 4, we obtain the following result immediately.\\
{\bf Corollary 1} The variations of coherence and
entanglement in Shor's algorithm are
\begin{equation}\label{eq36}
\Delta
C_{1,p}(\rho)=\left(r^{2}-1\right)^{\frac{1}{p}}-\left(Q-1\right)^{\frac{1}{p}},
\end{equation}
\begin{equation}\label{eq37}
\Delta C_{\alpha}(\rho)=\frac{1}{\alpha-1}\left[r^{2(1-\frac{1}{\alpha})}-Q^{1-\frac{1}{\alpha}}\right],
\end{equation}
\begin{equation}\label{eq38}
\Delta C_{g}(\rho)=\frac{1}{Q}-\frac{1}{r^{2}}
\end{equation}
and
\begin{equation}\label{eq39}
\Delta E_g(\rho)
=1-\frac{1}{r^{2}}\left(\sum_{a,s=0}^{r-1}\mathrm{e}^{-\frac{2\pi \mathrm{i}sa}{r}}\left(\frac{n-m_{a,s}}{n}\right)
^{\frac{n-m_{a,s}}{2}}
\left(\frac{m_{a,s}}{n}\right)^{\frac{m_{a,s}}{2}}\right)^{2},
\end{equation}
respectively.

Combining this with the inequality $Q \geq r^2$, we
can derive that $\Delta C_{1,p}$ $(\rho)<0$, $\Delta
C_{\alpha}(\rho)<0$ and $\Delta C_{g}(\rho)<0$. Corollary 1 reveals
a fundamental resource redistribution in Shor's algorithm, which
indicates that coherence is depleted throughout the computation. In
contrast, $\Delta E_g(\rho)>0$ suggests that entanglement is
actively generated. This dual behavior reflects a resource
conversion process, where coherence is consumed to fuel entanglement
production. Such dynamics highlight the interplay between coherence
and entanglement as key factors in the algorithm's efficiency. This
provides a critical insight for optimizing quantum algorithms: by
strategically managing coherence depletion and entanglement
generation, it may be possible to enhance computational efficiency
while minimizing resource costs.

\vskip0.1in

\noindent {\bf 5 Example }\\\hspace*{\fill}\\
In this section, we present an example to elucidate the characters of
coherence and entanglement in the state after the application of each basic
operator within Shor's algorithm.\\\hspace*{\fill}\\
\noindent {\bf Example 1} Let $N = 15$ and $x = 7$ that is coprime to $N$. The initial quantum state is $|0\rangle^{\otimes t}|1\rangle$, where the number $t$ of qubits in register $A$ is related to the precision. Here, we set $t=11$, which ensures that the probability of error is less than one quarter. The number $L$ of qubits in register $B$ needs to accommodate the binary representation of the integer $N$, that is, $L= 4$ and $n= 15$.\\
(i) Applying $t$ Hadamard gates to register $A$ yields the state $|\psi_{1}\rangle=\frac{1}{\sqrt{2^{11}}}\sum_{j=0}^{2^{11}-1}|j\rangle|1\rangle$. Let $\rho_{1}=|\psi_{1}\rangle\langle\psi_{1}|$, from Theorem 1, we have $C_{1,p}(\rho_{1})=\left(2^{11}-1\right)^{\frac{1}{p}}$, $C_{\alpha}(\rho_{1})=\frac{1}{\alpha-1}$ $\left(2^{11(1-\frac{1}{\alpha})}-1\right)$, $C_{g}(\rho_{1})\approx0.9995$ and $E_{g}(\rho_1)=0$.\\
(ii) The quantum state resulting from the application of the unitary
transformation $U$ is
$|\psi_{2}\rangle=\frac{1}{\sqrt{2^{11}}}\sum_{j=0}^{2^{11}-1}|j\rangle|7^{j}\mathrm{mod}
15\rangle$. $|7^{j}\mathrm{mod} 15\rangle$ takes on one of the four
states $|1\rangle$, $|7\rangle$, $|4\rangle$ and $|13\rangle$.
\begin{figure}[H]\centering
\subfigure[] {\begin{minipage}[figure1a]{0.49\linewidth}
\includegraphics[width=1.0\textwidth]{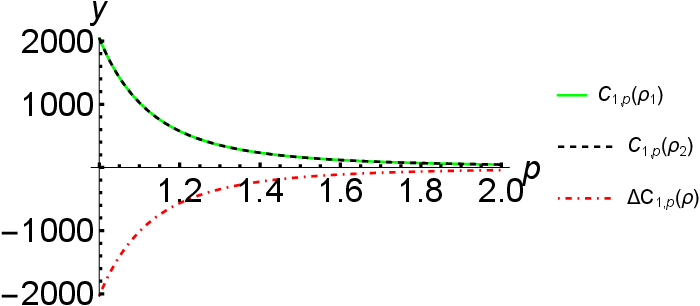}
\end{minipage}}
\subfigure[] {\begin{minipage}[figure1b]{0.49\linewidth}
\includegraphics[width=1.0\textwidth]{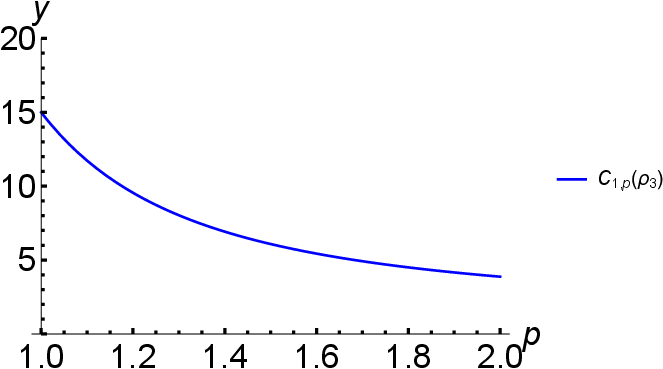}
\end{minipage}}
\subfigure[] {\begin{minipage}[figure2a]{0.49\linewidth}
\includegraphics[width=1.0\textwidth]{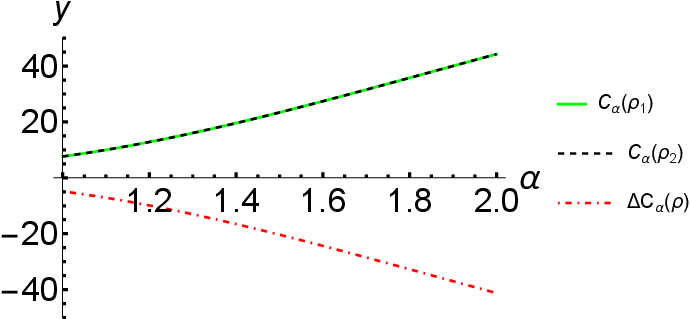}
\end{minipage}}
\subfigure[] {\begin{minipage}[figure2b]{0.49\linewidth}
\includegraphics[width=1.0\textwidth]{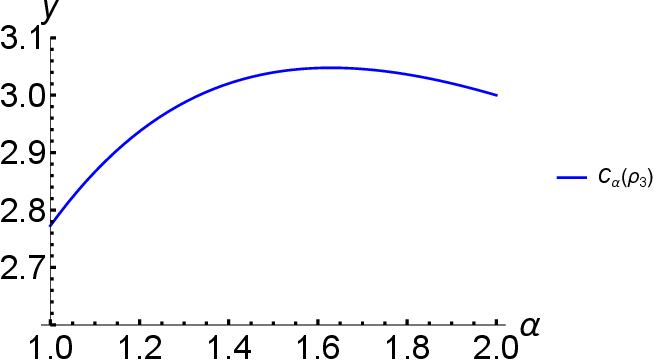}
\end{minipage}}
\subfigure[] {\begin{minipage}[figure3a]{0.49\linewidth}
\includegraphics[width=1.0\textwidth]{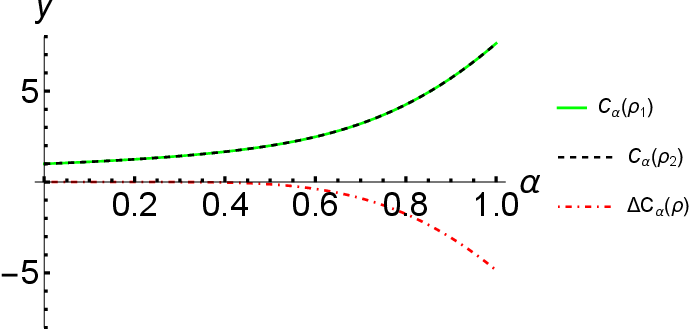}
\end{minipage}}
\subfigure[] {\begin{minipage}[figure3b]{0.49\linewidth}
\includegraphics[width=1.0\textwidth]{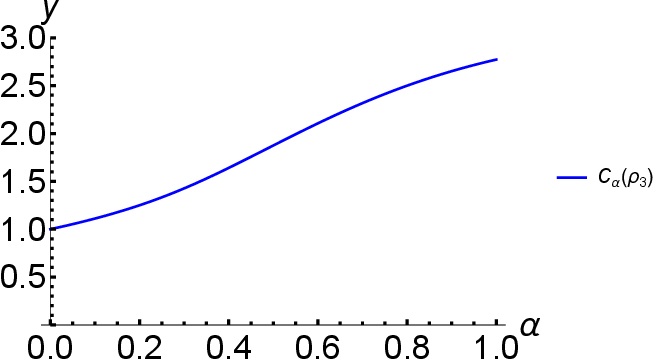}
\end{minipage}}
\caption{{Subfigures $\mathbf{a}$ and $\mathbf{b}$ ($\mathbf{c}$,
$\mathbf{d}$, $\mathbf{e}$ and $\mathbf{f}$) are for the case that
the coherence based on the $l_{1,p}$ norm (Tsallis relative $\alpha$
entropy). ($\mathbf{a}$, $\mathbf{b}$) The coherence with respect to $H$
(green), $U$ (black dashed), $F^{\dagger}$ (blue) and the variations
of coherence based on the $l_{1,p}$ norm (red dot-dashed).
($\mathbf{c}$, $\mathbf{d}$) The coherence with respect to $H$ (green),
$U$ (black dashed), $F^{\dagger}$ (blue) and the variations of
coherence based on the Tsallis relative $\alpha$ entropy (red
dot-dashed), where $\alpha\in(1,2]$. ($\mathbf{e}$, $\mathbf{f}$)
The coherence with respect to $H$ (green), $U$ (black dashed),
$F^{\dagger}$ (blue) and the variations of coherence based on the
Tsallis relative $\alpha$ entropy (red dot-dashed), where
$\alpha\in(0,1)$. \label{fig:Fig1}}}
\end{figure}
Let $\rho_{2}=|\psi_{2}\rangle\langle\psi_{2}|$, by Theorem 2, we have
$C_{1,p}(\rho_{2})=C_{1,p}(\rho_{1})$,
$C_{\alpha}(\rho_{2})=C_{\alpha}(\rho_{1})$,
$C_{g}(\rho_{2})=C_{g}(\rho_{1})\approx0.9995$ and
\begin{equation*}
E_g(\rho_2)=1-\frac{1}{2^{11}}\left(\sum_{a=0}^{3}\sum_{b=0}^{511}\left(\frac{15-n_{a,b}}{15}\right)
^{\frac{15-n_{a,b}}{2}}
\left(\frac{n_{a,b}}{15}\right)^{\frac{n_{a,b}}{2}}\right)^{2}\approx0.8445,
\end{equation*}
where $n_{a,b}$ is the Hamming weight of $|a+4b\rangle|7^{a} \mathrm{mod} 15\rangle$. \\
(iii) After applying the inverse Fourier transform $F^\dagger$ to
the register $A$, from Theorem 3, we can obtain
$C_{1,p}(\rho_{3})=15^{\frac{1}{p}}$,
$C_{\alpha}(\rho_{3})=\frac{1}{\alpha-1}\left(16^{1-\frac{1}{\alpha}}-1\right)$,
$C_{g}(\rho_{3})=0.9375$ and
\begin{equation*}
E_g(\rho_3)=1-\frac{1}{16}\left(\sum_{a=0}^{3}\sum_{s=0}^{3}\mathrm{e}^{\frac{-2\pi
\mathrm{i}sa}{4}}\left(\frac{15-m_{a,s}}{15}\right) ^{\frac{15-m_{a,s}}{2}}
\left(\frac{m_{a,s}}{15}\right)^{\frac{m_{a,s}}{2}}\right)^{2}\approx0.9876,
\end{equation*}
where $m_{a,s}$ is the Hamming weight of $|2^{9}s\rangle|7^{a} \mathrm{mod} 15\rangle$. \\
(iv) The variations of coherence based on the $l_{1,p}$ norm and the
Tsallis relative $\alpha$ entropy during the Shor's algorithm are
given by $\Delta
C_{1,p}(\rho)=15^{\frac{1}{p}}-\left(2^{11}-1\right)^{\frac{1}{p}}$
and
\begin{equation*}
\Delta
C_{\alpha}(\rho)=\frac{1}{\alpha-1}\left[16^{1-\frac{1}{\alpha}}
-2^{11(1-\frac{1}{\alpha})}\right],
\end{equation*}
respectively. Also, we have $\Delta C_{g}(\rho)=-0.062$ and $\Delta
E_{g}(\rho)=0.9876$.

In step (ii), if register $B$ measurement yields $|4\rangle$ from
$|7^{j}\mathrm{mod} 15\rangle$, then the state of register $A$
becomes
$\sqrt{\frac{4}{2^{t}}}[|2\rangle+|6\rangle+|10\rangle+|14\rangle+\ldots]$.
After applying $F^\dagger$, measurement will yield one of four
states: $|0\rangle$, $|512\rangle$, $|1024\rangle$ or
$|1536\rangle$. If $|1536\rangle$ is measured, then
$\frac{1536}{2048}=\frac{3}{4}$ gives $r=4$. Since
$7^{2}\mathrm{mod}15\neq\pm1$, we obtain $(7^{2}-1,15)=3$ and
$(7^{2}+1,15)=5$ as non-trivial factors of $N=15$.

For $H$, $U$ and $F^{\dagger}$ operations, coherence based on the $l_{1,p}$ norm decreases with increasing $p$, while its variation increases. Based on Tsallis relative $\alpha$ entropy, coherence for $H$ and $U$ increases with $\alpha$, while variation decreases within $\alpha\in(0,1)\cup(1,2]$. $F^{\dagger}$ exhibits more complex behavior: when $\alpha\in(1,2]$, coherence peaks at $\alpha^*\approx1.629$; when $\alpha\in(0,1)$, coherence consistently increases with $\alpha$ (see Fig. 1).

\vskip0.1in

\noindent {\bf 5 Conclusions and discussions}\\\hspace*{\fill}\\
We have investigated how coherence and entanglement impact on the
Shor's algorithm, by calculating how they change during each step
of the algorithm.
We have studied the variations of coherence and entanglement during Shor's algorithm and found that, irrespective of the employed coherence quantifier, coherence tends to deplete, while entanglement is generated throughout the process. Furthermore, the behavior of $F^{\dag}$ in Shor's algorithm can either consume or generate entanglement, depending on the order $r$, the Hamming weight, and the dimensions of each register.

Entanglement dynamics exhibit scale invariance for large $n$ in
Grover's search algorithm\cite{RBD}, implying that the geometric
entanglement does not depend on the number of qubits $n$ and is
solely determined by the ratio of iteration $k$ to the total number
of iterations. Here we show that the geometric entanglement is not
always scale invariant, but relies on the dimensions of each
register, the order $r$ and the Hamming weight of output state in
Shor's algorithm.

It has been observed that in Shor's algorithm, the coherence is
always depleted regardless of the coherence measures employed, which
is somewhat analogous to the role of quantum coherence in Grover's
search algorithm\cite{YLWZ1}. However, the coherence in HHL
algorithm maybe consumed or produced\cite{YLWZ2}, which is
determined by the explicit linear system of equations under
consideration, and the coherence could be produced or depleted
during the application of Simon's algorithm, depending on the
dimensionality of the system involved \cite{YLWZ3}.

These findings provide new theoretical support for understanding the efficiency of Shor's algorithm while offering novel perspectives on the role of quantum resources (such as coherence and entanglement) in quantum algorithms. Our approach can be employed to analyze coherence and entanglement dynamics in various quantum algorithms. Furthermore, this framework establishes theoretical foundations for optimizing quantum algorithm design, which may contribute significantly to advancing the field of quantum computing.

\vskip0.1in

\noindent


\subsubsection*{Declaration of competing interest}
\small {The authors declared no potential conflicts of interest with
respect to the research, authorship, and/or publication of this
article.}

\subsubsection*{Data availability}
\small {No new data were created or analysed in this study.}


\subsubsection*{Acknowledgements}
\small {This work was supported by National Natural Science
Foundation of China (Grant Nos. 12161056, 12075159, 12171044); Natural Science Foundation of Jiangxi Province (Grant No. 20232ACB211003);
Beijing Natural Science Foundation (Grant No. Z190005); the specific research fund of the Innovation Platform for Academicians of Hainan Province.}


\end{document}